\magnification=1200
\hoffset=.0cm
\voffset=.0cm
\baselineskip=.53cm plus .53mm minus .53mm

%
%
%
%
\def\ref#1{\lbrack#1\rbrack}
%
%
%
%
\input amssym.def
\input amssym.tex
%
%
\font\teneusm=eusm10                    
\font\seveneusm=eusm7                   
\font\fiveeusm=eusm5                 
%
%

%
%

%
%
\newfam\eusmfam
\textfont\eusmfam=\teneusm
\scriptfont\eusmfam=\seveneusm
\scriptscriptfont\eusmfam=\fiveeusm

\def\proclaim #1. #2\par{\medbreak{\bf #1.\enspace}{\it #2}\par\medbreak}
%
%
%
%
%

\def\hst1{\hskip 1pt}

%
%
%
%
%

\hbox to 16.5 truecm{September 1999   \hfil DFUB 99--18}
\hbox to 16.5 truecm{Version 1  \hfil hep-th/9909179}
\vskip2cm
\centerline{\bf Three Point Functions for a Class of Chiral Operators}
\centerline{\bf in Maximally Supersymmetric CFT at Large $N$} 
\vskip1cm
\centerline{by}
\vskip.5cm
\centerline{{\bf Fiorenzo Bastianelli}  and {\bf Roberto Zucchini}}
\centerline{\it Dipartimento di Fisica, Universit\`a degli Studi di Bologna}
\centerline{\it V. Irnerio 46, I-40126 Bologna, Italy}
\centerline{and}
\centerline{\it I. N. F. N., Sezione di Bologna}
\vskip1cm
\centerline{\bf Abstract} 
We present a calculation of three point functions for a class of 
chiral operators, including the primary ones,  
in $d=3$, ${\cal N}= 8$;  $d=6$, ${\cal N}=(2,0)$ and  
$d=4$, ${\cal N}= 4$ superconformal field theories at large $N$. 
These theories are related to the infrared world-volume descriptions 
of $N$ coincident M2, M5  and D3 branes, respectively.
The calculation is done in the framework of the AdS/CFT correspondence
and can be given a unified treatment employing a gravitational action
in arbitrary dimensions $D$, coupled to a $p+1$ form and suitably
compactified on ${\rm AdS}_{D-2-p} \times {\rm S}_{2+p}$.
The interesting cases are obtained setting 
$(D,p)$ to the values $(11,5)$, $(11,2)$ and $(10,3)$. 
\par\noindent
\vfill\eject
\par\vskip.6cm
\item{\bf 0.} {\bf Introduction}
\vskip.4cm
\par
The AdS/CFT correspondence [1--3] is a useful tool to study strongly 
coupled conformal field theories (CFT). 
It relates the generating functional for correlation functions
of gauge invariant operators in certain large $N$ CFT 
to the on-shell value of a corresponding dual supergravity 
action suitably compactified on AdS $\times$ M spaces.
When the dual pair is known, a tree-level computation on the
supergravity side gives a prediction on the otherwise 
inaccessible strongly coupled CFT.
The most studied case relates IIB supergravity on 
${\rm AdS}_5 \times {\rm S}_5$ to $d=4$, ${\cal N} =4$
large $N$ $SU(N)$ super Yang--Mills (SYM) theory, where a
variety of correlation functions have already been computed 
using the correspondence (see the review [4] also for a list of references).
In  particular, three point functions for a set of chiral primary 
operators (CPO), namely those corresponding to the so-called single 
trace operators, have been computed in [5].
Other three point functions involving the CPOs have been computed more
recently in [6,7], in part with the aim of addressing eventually 
the more complicated case of their four point functions [8].

The $d=4$, ${\cal N} =4$ SYM theory, which can be thought of
as describing the low energy dynamics of D3 branes in
type IIB superstring theory,
is not the only example with 16 supersymmetric charges.
Other dual pairs involving maximally supersymmetric CFT
have been identified in [1], namely 
``11D supergravity on  ${\rm AdS}_7 \times {\rm S}_4$ /
$d=6$, ${\cal N}=(2,0)$ large $N$ SCFT'' and
``11D supergravity on  ${\rm AdS}_4 \times {\rm S}_7$ / 
$d=3$, ${\cal N}= 8$   large $N$ SCFT''.
These pairs describe the low energy physics of $N$ coincident M5 and M2
branes of M-theory, respectively. In this respect,
the AdS/CFT correspondence has been used in [9]
to identify CPO three point functions in the 
$d=6$, ${\cal N}=(2,0)$  SCFT case, while the present authors
produced in [10] the set of CPO three point functions for both cases.
In this last reference, we could treat simultaneously both cases by 
employing a  gravitational action in arbitrary dimensions $D$, 
coupled to a $1+p$ form, suitably
compactified on ${\rm AdS}_{D-2-p} \times {\rm S}_{2+p}$.
Setting at the end $(D,p)$ to the values $(11,2)$  and $(11,5)$ 
allowed us to recover the M5 and M2 cases, respectively.
We also noticed, with some surprise, that the general model with 
$(D,p)=(10,3)$ was able to reproduce the super Yang--Mills case 
worked out in [5].

As promised in [10], in this paper we present more details on the 
calculation leading to the  CPO three point functions
for all of these maximally supersymmetric CFTs.
In addition, we extend it to include another class
of scalar chiral operators sitting in the same short 
supermultiplets identified by the chiral primaries.
Including them gives us the chance of producing quite large a
class of three point functions with minor efforts.
Note that obtaining the additional three point functions from the CPO ones
by using the superconformal algebra seems possible in principle, 
but may be rather cumbersome.
The sources for the CPOs are given according to the AdS/CFT correspondence
by the Kaluza--Klein tower of scalar fields with lowest mass,
to be denoted by $s_I$, while the extra primary operators we wish to 
consider are scalar operators which couple to another Kaluza--Klein
tower of scalars to be denoted by $t_I$.
These $s_I$ - $t_I$ fields are identified as mass eigenstates 
contained in the perturbations of the 
trace of the metric on ${\rm S}_{2+p}$
and of the $1+p$ form on ${\rm S}_{2+p}$.
Also, we show why our general model is expected to reproduce
the super Yang--Mills case by setting $(D,p)=(10,3)$.
This gives us the opportunity of computing
a new class of three point functions for the latter theory.

The paper is organized as follows. In sect. 1 we show why and how
all these various cases of maximally supersymmetric CFT can be treated
in a unified scheme which employs a general gravitational 
action. In sect. 2 we present the calculation identifying the bulk
cubic couplings of those physical fluctuations which couple to 
the primary operators of our interest.
In sect. 3 we use the couplings to compute the normalized three
point functions for the given set of primary operators.
Finally, in sect. 4 we present our conclusions and leave appendix A
to set some notations for the scalar spherical harmonics
on the sphere ${\rm S}_n$.
\par\vskip.6cm
\item{\bf 1.} {\bf Justification of the method}
\vskip.4cm
\par
We consider first the case of branes in M-theory.
According to the AdS/CFT correspondence principles [1--4], 
the low energy world volume conformal 
field theory of $N$ coincident M5 (M2) branes at large $N$ is described by 
$D=11$ supergravity compactified on ${\rm AdS}_7\times{\rm S}_4$ 
(${\rm AdS}_4\times{\rm S}_7$) [11--13].
The conformal operators of the CFT on the AdS boundary are related by duality
in a precise way to the operators of the bulk AdS field theory representing 
the fluctuations of the supergravity fields around a maximally supersymmetric 
Freund--Rubin background [14]. As explained in the introduction, the aim of 
this paper is the computation of a class of bosonic three point functions
of these CFTs. Thus, we need to consider only a certain subset of bosonic 
fluctuations described below.
 
The action of the bosonic sector in the usual 3 form formulation of $D=11$ 
supergravity on $M_{11}$ is given by
$$
I={1\over 4\kappa^2}\int_{M_{11}}\Big[R(g)*_g1-F_4\wedge *_gF_4
+\hbox{$2^{1\over 2}\over 3$}A_3\wedge F_4\wedge F_4\Big].
\eqno(1.1)
$$
Here, $R(g)$ is the Ricci scalar of the metric $g$ and
$F_4=dA_3$ is the 4 form field strength of the 3 form field $A_3$.
 
For the M5 theory, $M_{11}={\rm AdS}_7\times{\rm S}_4$. The Freund--Rubin 
background $\bar g$, $\bar A_3$ is such that $\bar g$ is factorized 
and $\bar F_4=\bar F_{(0,4)}$
\footnote{}{}
\footnote{${}^1$}{Consider the $D$ dimensional space time
$M_D={\rm AdS}_{D-2-p}\times{\rm S}_{2+p}$. 
We say that a metric $g$ on $M_D$ is factorized if $g$ has the block 
structure $g=g'\oplus g''$, where $\bar g'$, $\bar g''$ are metrics
on ${\rm AdS}_{D-2-p}$, ${\rm S}_{2+p}$, respectively.
We denote form degree on $M_D$ by a subfix, 
e. g. $\omega_r$ is a $r$ form
on $M_D$. Similarly, we denote form degree on the factors ${\rm AdS}_{D-2-p}$, 
${\rm S}_{2+p}$ by a pair of suffixes, e. g. $\omega_{(r,s)}$ denotes a 
$r+s$ form on $M_D$ that is a $r$ form on ${\rm AdS}_{D-2-p}$ and a $s$ form 
on ${\rm S}_{2+p}$.}. 
The fluctuations of $g$ and $A_3$ around the background, relevant in 
the analysis that follows, are such that $g$ remains factorized and 
$F_4=\bar F_{(0,4)}+da_{(0,3)}$. By dimensional reasons,
for such fluctuations the integrand of the Chern--Simons term vanishes 
identically. Thus, we may use the truncated action
$$
I={1\over 4\kappa^2}\int_{{\rm AdS}_7\times{\rm S}_4}
\Big[R(g)*_g1-F_4\wedge *_gF_4\Big],
\eqno(1.2)
$$
$$
g=g'\oplus g'',\quad F_4=\bar F_{(0,4)}+da_{(0,3)}.
\eqno(1.3)
$$

For the M2 theory, $M_{11}={\rm AdS}_4\times{\rm S}_7$. The Freund--Rubin 
background $\bar g$, $\bar A_3$ is such that $\bar g$ is factorized
and $\bar F_4=\bar F_{(4,0)}$. The relevant fluctuations of $g$ and 
$A_3$ around the background are such that $g$ remains factorized and 
$F_4=\bar F_{(4,0)}+da_{(3,0)}$. 
The integrand of the Chern--Simons term vanishes identically
for such fluctuations in this case as well by dimensional reasons.
Thus, we may use the action
$$
I={1\over 4\kappa^2}\int_{{\rm AdS}_4\times{\rm S}_7}
\Big[R(g)*_g1-F_4\wedge *_gF_4\Big],
\eqno(1.4)
$$
$$
g=g'\oplus g'',\quad F_4=\bar F_{(4,0)}+da_{(3,0)}.
\eqno(1.5)
$$
Superficially, this seems to parallel the M5 case closely.
However, a closer inspection reveals that the relevant scalar fluctuation
contained in $a_{(3,0)}$ comes about as the solution of the constraint
$$
d*_{\bar g'}a_{(3,0)}=0
\eqno(1.6)
$$ 
entailed by gauge fixing at quadratic level, which is difficult to 
implement in an off--shell fashion. This problem can be solved by means of a
standard dualization trick. On notes that the action 
$\int(-dA_3\wedge *_gdA_3)$ is the reduction of the more general action
$\int(-F_4\wedge *_gF_4+2F_7\wedge (F_4-dA_3))$ upon substituting the 
$F_7$ field equation $F_4-dA_3=0$. Substituting first instead the 
$A_3$ and $F_4$ field equations $dF_7=0$, $F_7=*_gF_4$, one gets the 
equivalent dual action $\int(-dA_6\wedge *_gdA_6)$, where $A_6$ is the 6 form
field solving the Bianchi identity $dF_7=0$. In the dual formulation,
the Freund--Rubin background $\bar g$, $\bar A_6$ is 
such that $\bar g$ is again factorized and 
$\bar F_7=\bar F_{(0,7)}$. The relevant fluctuations of $g$ and $A_6$ 
around the background are such that $g$ remains factorized and 
$F_7=\bar F_{(0,7)}+da_{(0,6)}$. To summarize, the action is 
$$
I={1\over 4\kappa^2}\int_{{\rm AdS}_4\times{\rm S}_7}
\Big[R(g)*_g1-F_7\wedge *_gF_7\Big],
\eqno(1.7)
$$
$$
g=g'\oplus g'',\quad F_7=\bar F_{(0,7)}+da_{(0,6)}.
\eqno(1.8)
$$

We now turn to the case of D3 branes in type IIB superstring theory.
According to the AdS/CFT correspondence, the low energy world volume 
conformal field theory of a large number of coincident D3 branes 
is described by $D=10$ type IIB supergravity compactified on 
${\rm AdS}_5\times{\rm S}_5$. Again, the conformal operators of the CFT 
on the AdS boundary are related by duality in a precise way to the operators 
of the bulk AdS field theory representing the fluctuations of the IIB 
supergravity fields around a maximally supersymmetric selfdual Freund--Rubin 
background [15]. To compute the set of bosonic three point functions
we are interested in, we need to consider only a 
certain subset of bosonic fluctuations described below.
We may try to proceed as we did above dealing with M theory and begin our 
discussion from the fully covariant action of the bosonic sector of type IIB
supergravity worked out in ref. [16], but this does not seem to bring us 
that far. Thus, we follow a different path which we are now going to explain.

For the D3 brane, space time is $M_{10}={\rm AdS}_5\times{\rm S}_5$.
The relevant fields are the metric $g$ and the Ramond--Ramond 4 form
field $A_4$ with selfdual field strength $F^{sd}_5=dA_4$
$$
F^{sd}_5=*_gF^{sd}_5.
\eqno(1.9)
$$
Further, Einstein's field equations hold 
$$
R(g)_{ij}*_g1=F^{sd}_5{}_i \wedge *_g F^{sd}_5{}_j
\eqno(1.10)
$$
where $R(g)_{ij}$ is the Ricci tensor and $F^{sd}_5{}_i={1\over 4!}
F^{sd}_{ijklm}dx^j\wedge dx^k\wedge dx^l\wedge dx^m$.

The selfdual Freund--Rubin background $\bar g$, $\bar A_4$ 
is such that $\bar g$ is factorized and $\bar F^{sd}_5
=2^{-{1\over 2}}(\bar F_{(0,5)}+*_{\bar g}\bar F_{(0,5)})$.
The relevant fluctuations of $g$ and $A_4$ around the background
are such that $g$ is factorized, as usual, and $F^{sd}_5
=\bar F^{sd}_5+da_4$, where $a_4=2^{-{1\over 2}}(a_{(4,0)}+a_{(0,4)})$. 
The selfduality equations relate the fluctuations $a_{(4,0)}$, $a_{(0,4)}$ 
as
$$
\eqalignno{
&*_g(d_{(1,0)}a_{(0,4)})-d_{(0,1)}a_{(4,0)}=0,&(1.11a)\cr
&*_g(d_{(0,1)}a_{(0,4)})-d_{(1,0)}a_{(4,0)}
+(*_g-*_{\bar g})\bar F_{(0,5)}=0.&(1.11b)\cr}
$$
Therefore, $a_{(4,0)}$ is not independent of $a_{(0,4)}$ as eq. (1.11a)
allows in principle to express $a_{(4,0)}$ in terms of $a_{(0,4)}$.
Taking the selfduality equations into account, one can check that 
Einstein's equation can be written in terms of $a_{(0,4)}$ only
$$
\eqalignno{
R(g)_{ij}*_g1&=
(\bar F_{(0,5)}+da_{(0,4)}){}_i\wedge *_g(\bar F_{(0,5)}+da_{(0,4)}){}_j
&\cr
&\hphantom{=~}-\hbox{$1\over 2$}g_{ij}(\bar F_{(0,5)}+da_{(0,4)})\wedge *_g
(\bar F_{(0,5)}+da_{(0,4)}).&(1.12)\cr}
$$

Let us examine the above field equations from another point of view.
We consider an action of the form
$$
I=\int_{{\rm AdS}_5\times{\rm S}_5}\Big[R(g)*_g1-F_5\wedge *_gF_5\Big],
\eqno(1.13)
$$
where $g$ is a metric and $F_5=dA_4$ is a 5 form field strength,
not necessarily selfdual. We pick a Freund--Rubin like background 
$\bar g$, $\bar A_4$ where $\bar g$ is factorized as usual 
and $\bar F_5=\bar F_{(0,5)}$.
We restrict to fluctuations for which $g$ remains factorized 
and $F_5=\bar F_{(0,5)}+da_{(0,4)}$.
The field equations deduced from this action for the $a_{(0,4)}$ 
fluctuations are
$$
\eqalignno{
&d_{(0,1)}\big(*_g(d_{(1,0)}a_{(0,4)})\big)=0,&(1.14a)\cr
&d_{(0,1)}\big(*_g(d_{(0,1)}a_{(0,4)})
+(*_g-*_{\bar g})\bar F_{(0,5)}\big)
+d_{(1,0)}\big(*_g(d_{(1,0)}a_{(0,4)})\big)=0.&(1.14b)\cr}
$$
From (1.14a), recalling that $H^1_{\rm de Rham}({\rm S}_5)=0$, one has 
$$
*_g(d_{(1,0)}a_{(0,4)})=d_{(0,1)}a_{(4,0)}
\eqno(1.15)
$$
for some 4 form $a_{(4,0)}$. Substituting this relation in (1.14b),
one gets
$$
d_{(0,1)}\big(*_g(d_{(0,1)}a_{(0,4)})-d_{(1,0)}a_{(4,0)}
+(*_g-*_{\bar g})\bar F_{(0,5)}\big)=0.
\eqno(1.16)
$$
Using that $H^0_{\rm de Rham}({\rm S}_5)=\Bbb R$, this equation can 
be integrated once, yielding
$$
*_g(d_{(0,1)}a_{(0,4)})-d_{(1,0)}a_{(4,0)}
+(*_g-*_{\bar g})\bar F_{(0,5)}=\omega_{(5,0)},
\eqno(1.17)
$$
where the 5 form $\omega_{(5,0)}$ satisfies
$$ 
d_{(0,1)}\omega_{(5,0)}=0,
\eqno(1.18)
$$
but is otherwise arbitrary. As $H^5_{\rm de Rham}({\rm AdS}_5)
=0$, there is a 4 form $\omega_{(4,0)}$ such that
$$
\omega_{(5,0)}=d_{(1,0)}\omega_{(4,0)},\quad d_{(0,1)}\omega_{(4,0)}=0.
\eqno(1.19)
$$
Now, from (1.15), it is evident that $a_{(4,0)}$ is not uniquely defined 
and might be redefined into $a_{(4,0)}-\omega_{(4,0)}$. 
By doing so, we can assume
that 
$$
\omega_{(5,0)}=0
\eqno(1.20)
$$
in eq. (1.17). We could have arrived to eq. (1.20) more directly by noting 
that we need to consider only small perturbations 
around the chosen background. Taking the limit of vanishing
perturbations in eq. (1.17) and requiring that the background 
satisfies the resulting equation of motion fixes $\omega_{(5,0)}=0$.
In conclusion, eqs. (1.14a), (1.14b) have the same content as eqs. 
(1.11a), (1.11b). The Einstein's field equations following from (1.13)
are the same as (1.12).
Thus, by the above reasoning, the selfdual dynamics 
of the relevant bosonic type IIB fluctuations is reduced to that of the 
the field $g$ and $a_{(0,4)}$ with action 
$$
I={1\over 4\kappa^2}\int_{{\rm AdS}_5\times{\rm S}_5}
\Big[R(g)*_g1-F_5\wedge *_gF_5\Big],
\eqno(1.21)
$$
$$
g=g'\oplus g'',\quad F_5=\bar F_{(0,5)}+da_{(0,4)}.
\eqno(1.22)
$$

The upshot of the above discussion is that all three models described 
above may be treated on the same footing as follows, in spite of their 
different physical content.

Space time $M_D$ is ${\rm AdS}_{D-2-p}\times {\rm S}_{2+p}$.
The relevant fields are the metric $g$ and the $1+p$ form field
$A_{1+p}$. The Freund--Rubin background $\bar g$, $\bar A_{1+p}$ is such 
that $\bar g$ is factorized and $\bar F_{2+p}=\bar F_{(0,2+p)}$. 
The relevant bosonic fluctuations of 
$g$ and $A_{1+p}$ around the background are such that $g$ remains 
factorized and $F_{2+p}=\bar F_{(0,2+p)}+da_{(0,1+p)}$.
The action effectively describing the dynamics of the fluctuations is 
$$
I={1\over 4\kappa^2}\int_{{\rm AdS}_{D-2-p}\times{\rm S}_{2+p}}
\Big[R(g)*_g1-F_{2+p}\wedge *_gF_{2+p}\Big],
\eqno(1.23)
$$
$$
g=g'\oplus g'',\quad F_{2+p}=\bar F_{(0,2+p)}+da_{(0,1+p)}.
\eqno(1.24)
$$
The physically important cases are those for which 
$(D,p)=(11,2),~(11,5),~(10,3)$. 

\par\vskip.6cm
\item{\bf 2.} {\bf Implementation of the method}
\vskip.4cm
\par
We are now going to concretely
implement the method described in the former section.
Our starting point is the general action in eq. (1.23).
One can easily verify that the field equations  
admit the standard ${\rm AdS}_{D-2-p} \times {\rm S}_{2+p}$
solution $\bar g_{ij}$,
$\bar A_{i_1\cdots i_{1+p}}$ generated by the Freund Rubin ansatz [14].
For this, the only non vanishing components of the Riemann tensor 
and field strength are given by
$$
\eqalignno{ 
\bar R_{\kappa\lambda\mu\nu}
&=-a_1(\bar g_{\kappa\mu}\bar g_{\lambda\nu}
-\bar g_{\kappa\nu}\bar g_{\lambda\mu}),  
\quad\quad a_1={(1+p)\over{(D-2)(D-3-p)}}e^2,&(2.1a)\cr
\bar R_{\alpha\beta\gamma\delta}
&=a_2(\bar g_{\alpha\gamma}\bar g_{\beta\delta}
-\bar g_{\alpha\delta}\bar g_{\beta\gamma}),
\quad\quad a_2={(D-3-p)\over (D-2)(1+p)}e^2,&(2.1b)\cr
\bar F_{\alpha_1\cdots\alpha_{2+p}}
&=e\bar \epsilon_{\alpha_1\cdots\alpha_{2+p}},
\quad\quad{\rm other}~{\bar F}=0, 
\vphantom{(D-3-p)\over (D-2)(1+p)}  &(2.2)\cr}
$$
where $\bar \epsilon_{\alpha_1\cdots\alpha_{2+p}}$ denotes the standard
volume form on the unit sphere and $e$ is an arbitrary mass scale 
parametrizing the compactification
\footnote{}{}
\footnote{${}^2$}{In this paper, we adopt the following conventions. 
Latin lower case letters $i,j,k,l,\ldots$ denote $M_D$ indices.
Late Greek lower case letters $\kappa,\lambda,\mu,\nu\dots$
denote ${\rm AdS}_{D-2-p}$ indices.
Early Greek lower case letters $\alpha,\beta,\gamma,\delta,\ldots$
denote ${\rm S}_{2+p}$ indices.}.

We expand the action in fluctuations around the
background $\bar g_{ij}$, $\bar A_{i_1\cdots i_{1+p}}$.
We parametrize the fluctuations $\delta g_{ij}$, $\delta A_{i_1\cdots i_{1+p}}$
of the fields $g_{ij}$, $A_{i_1\cdots i_{1+p}}$ around the background 
as in [11]
$$
\eqalignno{
\delta g_{\kappa\lambda}&=h_{\kappa\lambda}
-\hbox{$1\over D-4-p$}\bar g_{\kappa\lambda}\pi,\vphantom{\Big[}&(2.3a)\cr
\delta g_{\alpha\kappa}&=k_{\alpha\kappa}+\bar\nabla_\alpha l_\kappa,
\vphantom{\Big[}&\cr
\bar\nabla^\gamma k_{\gamma\kappa}&=0,\vphantom{\Big[}&(2.3b)\cr
\delta g_{\alpha\beta}&=m_{\alpha\beta}+\bar\nabla_\alpha n_\beta
+\bar\nabla_\beta n_\alpha+(\bar\nabla_\alpha\bar\nabla_\beta
-\hbox{$1\over 2+p$}\bar g_{\alpha\beta}\bar\nabla^\gamma\bar\nabla_\gamma)q
+\hbox{$1\over 2+p$}\bar g_{\alpha\beta}\pi,\vphantom{\Big[}&\cr
m^\gamma{}_\gamma&=0, \quad \bar\nabla^\gamma m_{\gamma\alpha}=0,\quad
\bar\nabla^\gamma n_\gamma=0,\vphantom{\Big[}&(2.3c)\cr}
$$
$$
\eqalignno{
\delta A_{\alpha_1\cdots\alpha_{1+p}}&=
(1+p)\bar\nabla_{[\alpha_1}a_{\alpha_2\cdots\alpha_{1+p}]}
+\bar\epsilon_{\alpha_1\cdots\alpha_{1+p}}{}^\gamma\bar\nabla_\gamma b,
\vphantom{\Big[}&\cr
\bar\nabla^\gamma a_{\gamma\alpha_3\cdots\alpha_{1+p}}&=0.&(2.4)\cr}
$$
Fluctuations of the other components of $A_{i_1\cdots i_{1+p}}$ 
can be disregarded as they are independent from the ones we are interested in,
and we can set them directly to zero.
Before describing the calculation, we note that one could partially fix the 
gauge by eliminating those gauge invariances that do not correspond
to the usual reparametrization and form gauge invariances
from the  ${\rm AdS}_{D-2-p}$ perspective. This could be done
by imposing
$$
\bar\nabla^\beta\big(\delta g_{\beta\alpha}
-\hbox{$1\over 2+p$}\bar g_{\beta\alpha}\delta g^\gamma{}_\gamma\big)=0,\quad
\bar\nabla^\beta\delta g_{\beta\kappa}=0;
\eqno(2.5a)-(2.5b)
$$
$$
\bar\nabla^\beta\delta A_{i_1\cdots i_p\beta}=0,
\eqno(2.6)
$$
as shown in [11]. These conditions imply in particular that 
$\bar\nabla_\alpha l_\kappa=0$,
$\bar\nabla_\alpha n_\beta+\bar\nabla_\beta n_\alpha=0$, 
$(\bar\nabla_\alpha\bar\nabla_\beta-\hbox{$1\over 2+p$}
\bar g_{\alpha\beta}\bar\nabla^\gamma\bar\nabla_\gamma)q=0$ 
and $a_{\alpha_1\cdots\alpha_p}=0$. 
So, upon gauge fixing, the field equations of the fields $l_\kappa$,
$n_\alpha$, $q$ and $a_{\alpha_1\cdots\alpha_p}$  
must be enforced by hand as constraints.
However, in the sequel we will not need to proceed this way. Rather,
we will identify the action describing the dynamics of the physical 
scalar fields contained in the $\pi$-$b$ fluctuations by taking care
of the constraints associated with the various
gauge invariances using field redefinitions.

The quadratic and cubic actions of the fluctuation fields are given by
$$
\eqalignno{
I_{[2]}&={1\over 4\kappa^2}
\int_{{\rm AdS}_{D-2-p}}d^{D-2-p}y (-\bar g_{D-2-p})^{1\over 2}
\int_{{\rm S}_{2+p}}d^{2+p}x (\bar g_{2+p})^{1\over 2}&\cr
&\bigg\{\hbox{$1\over 2$}\bar\nabla^\kappa h_{\lambda\mu}\bar\nabla^\lambda
h_\kappa{}^\mu\vphantom{\bigg\{}
-\hbox{$1\over 2$}\bar\nabla^\kappa h_{\kappa\lambda}\bar\nabla^\lambda 
h^\mu{}_\mu
+\hbox{$1\over 4$}\bar\nabla^\kappa h^\lambda{}_\lambda\bar\nabla_\kappa 
h^\mu{}_\mu
-\hbox{$1\over 4$}\bar\nabla^\kappa h^{\lambda\mu}\bar\nabla_\kappa 
h_{\lambda\mu}\vphantom{\bigg\{}&\cr
&+\hbox{$1\over 4$}\bar\nabla^\alpha h^\kappa{}_\kappa\bar\nabla_\alpha 
h^\lambda{}_\lambda
-\hbox{$1\over 4$}\bar\nabla^\alpha h^{\kappa\lambda}\bar\nabla_\alpha 
h_{\kappa\lambda}
+{1+p\over 2(D-2)}e^2\bigg[
\hbox{$1\over 2$}h^\kappa{}_\kappa h^\lambda{}_\lambda
-h^{\kappa\lambda}h_{\kappa\lambda}\bigg]
\vphantom{\bigg\{}&\cr
&-{D-2\over 4(2+p)(D-4-p)}\bar\nabla^\kappa\pi\bar\nabla_\kappa\pi
+{(D-2)(4+(-D+6)p+p^2)\over 4(2+p)^2(D-4-p)^2}
\bar\nabla^\alpha\pi\bar\nabla_\alpha\pi\vphantom{\bigg\{}&\cr
&-{(1+p)(D-3-p)\over 2(2+p)(D-4-p)}e^2\pi\pi
+ {D-2\over 2(2+p)(D-4-p)}
h^\kappa{}_\kappa\bar\nabla^\alpha\bar\nabla_\alpha\pi
\vphantom{\bigg\{}&\cr
& -\bar\nabla^\kappa\bar\nabla^\alpha b\bar\nabla_\kappa\bar\nabla_\alpha b
-\bar\nabla^\alpha\bar\nabla_\alpha b\bar\nabla^\beta\bar\nabla_\beta b
-2(-1)^p{(D-3-p)\over(D-4-p)}e\pi\bar\nabla^\alpha\bar\nabla_\alpha b
\vphantom{\bigg\{}&\cr
&+(-1)^p eh^\kappa{}_\kappa\bar\nabla^\alpha\bar\nabla_\alpha b
+\ldots\bigg\},&(2.7)\cr}
$$
and
$$
\eqalignno{
I_{[3]}&={1\over 4\kappa^2}
\int_{{\rm AdS}_{D-2-p}}d^{D-2-p}y (-\bar g_{D-2-p})^{1\over 2}
\int_{{\rm S}_{2+p}}d^{2+p}x (\bar g_{2+p})^{1\over 2}&\cr
&\bigg\{\hbox{$1\over 4$}h^\nu{}_\nu\bigg[
\bar\nabla^\kappa h^{\lambda\mu}\bar\nabla_\lambda h_{\kappa\mu}
-\bar\nabla^\kappa h_{\kappa\lambda}\bar\nabla^\lambda h^\mu{}_\mu
+\hbox{$1\over 2$}\bar\nabla^\kappa h^\lambda{}_\lambda\bar\nabla_\kappa 
h^\mu{}_\mu
-\hbox{$1\over 2$}\bar\nabla^\kappa h^{\lambda\mu}\bar\nabla_\kappa  
h_{\lambda\mu}\bigg]\vphantom{\bigg\{}&\cr
&+\hbox{$1\over 4$}h^{\kappa\lambda}h_{\kappa\lambda}\bigg[
\bar\nabla^\mu\bar\nabla_\mu h^\nu{}_\nu
-\bar\nabla^\mu \bar\nabla^\nu h_{\mu\nu}\bigg]
-\hbox{$1\over 2$}h^{\kappa\lambda}\bigg[
2\bar\nabla_\kappa h^{\mu\nu}\bar\nabla_\mu h_{\lambda\nu}
\vphantom{\bigg\{}&\cr
&+\bar\nabla^\mu h_{\kappa\nu}\bar\nabla^\nu h_{\lambda\mu}
-\bar\nabla_\kappa h_{\lambda\mu}\bar\nabla^\mu h^\nu{}_\nu
-\bar\nabla_\kappa h^\mu{}_\mu\bar\nabla^\nu h_{\nu\lambda}
-\bar\nabla^\mu h_\kappa{}^\nu\bar\nabla_\mu h_{\lambda\nu}
\vphantom{\bigg\{}&\cr
&+\hbox{$1\over 2$}\bar\nabla_\kappa h^\mu{}_\mu
\bar\nabla_\lambda h^\nu{}_\nu
-\hbox{$1\over 2$}\bar\nabla_\kappa h^{\mu\nu}
\bar\nabla_\lambda h_{\mu\nu}\bigg]
+\hbox{$1\over 4$}\bigg[h^{\kappa\lambda}h_{\kappa\lambda}
-\hbox{$1\over 4$}h^\kappa{}_\kappa h^\lambda{}_\lambda\bigg]
\bar\nabla^\alpha\bar\nabla_\alpha h^\mu{}_\mu\vphantom{\bigg\{}&\cr
&+\hbox{$1\over 2$}h^{\kappa\lambda}\bar\nabla^\alpha h_\kappa{}^\mu
\bar\nabla_\alpha h_{\lambda\mu}
-\hbox{$1\over 8$}h^\kappa{}_\kappa\bar\nabla^\alpha h^{\lambda\mu}
\bar\nabla_\alpha h_{\lambda\mu}
\vphantom{\bigg\{}&\cr
&+{1+p\over D-2}e^2\bigg[
\hbox{$1\over 12$}h^\kappa{}_\kappa h^\lambda{}_\lambda h^\mu{}_\mu
-\hbox{$1\over 2$}h^\kappa{}_\kappa h^\lambda{}_\mu h^\mu{}_\lambda
+\hbox{$2\over 3$}h^\kappa{}_\lambda h^\lambda{}_\mu h^\mu{}_\kappa
\bigg]\vphantom{\bigg\{}&\cr
&+{1\over 2(D-4-p)}\pi\bigg[
h^{\kappa\lambda}\bar\nabla^\mu\bar\nabla_\mu h_{\kappa\lambda}
-h^\kappa{}_\kappa\bar\nabla^\mu\bar\nabla_\mu h^\lambda{}_\lambda
+h^\mu{}_\mu\bar\nabla^\kappa\bar\nabla^\lambda h_{\kappa\lambda}
\vphantom{\bigg\{}&\cr
&+h_{\kappa\lambda} \bar\nabla^\kappa\bar\nabla^\lambda h^\mu{}_\mu
-2 h^{\kappa\lambda} \bar\nabla^\mu\bar\nabla_\kappa h_{\lambda\mu}\bigg]
+{D-6-2p\over 4(2+p)(D-4-p)}
\pi h^\kappa{}_\kappa\bar\nabla^\alpha\bar\nabla_\alpha h^\lambda{}_\lambda
\vphantom{\bigg\{}&\cr
&-{1\over 2(2+p)}
\pi h^{\kappa\lambda}\bar\nabla^\alpha\bar\nabla_\alpha h_{\kappa\lambda}
-{D-2\over 4(2+p)(D-4-p)}
\pi \bar\nabla^\alpha h^{\kappa\lambda}\bar\nabla_\alpha h_{\kappa\lambda}
\vphantom{\bigg\{}&\cr
&+{1+p\over (D-2)(D-4-p)}e^2\pi\bigg[
\hbox{$1\over 2$}h^\kappa{}_\kappa h^\lambda{}_\lambda 
-h^\kappa{}_\lambda h^\lambda{}_\kappa \bigg]
\vphantom{\bigg\{}&\cr
&-{D-2\over 2(2+p)(D-4-p)}\bigg[h^{\kappa\lambda}
\bigg(\pi\bar\nabla_\kappa\bar\nabla_\lambda\pi
+\hbox{$1\over 2$}\bar\nabla_\kappa\pi\bar\nabla_\lambda\pi\bigg)
-h^\kappa{}_\kappa\bigg(\pi\bar\nabla^\lambda\bar\nabla_\lambda\pi
\vphantom{\bigg\{}&\cr
&+\hbox{$3\over 4$}\bar\nabla^\lambda\pi\bar\nabla_\lambda\pi\bigg)\bigg]
+{(D-2)(4+(3D-10)p-3p^2)\over 8(2+p)^2(D-4-p)^2}
h^\kappa{}_\kappa \bar\nabla^\alpha \pi\bar\nabla_\alpha \pi
\vphantom{\bigg\{}&\cr
&+{(D-2)(2+(D-3)p-p^2)\over 2(2+p)^2(D-4-p)^2}
h^\kappa{}_\kappa \pi\bar\nabla^\alpha\bar\nabla_\alpha \pi
-{(D-2-p)(1+p)\over 4(2+p)(D-4-p)}e^2 h^\kappa{}_\kappa\pi\pi
\vphantom{\bigg\{}&\cr
&+{D-2\over 2(2+p)^2(D-4-p)}\pi\bar\nabla^\kappa\pi\bar\nabla_\kappa \pi
\vphantom{\bigg\{}&\cr
&+{(D-2)(4D^2-28D+56+(5D^2-40D+84)p-(7D-26)p^2+2p^3)\over 4(2+p)^3(D-4-p)^3}
\pi\bar\nabla^\alpha \pi\bar\nabla_\alpha \pi
\vphantom{\bigg\{}&\cr
&+{(1+p)(D-3-p)(6D-20+(D-8)p-p^2)\over 6(2+p)^2(D-4-p)^2}\pi\pi\pi
\vphantom{\bigg\{}&\cr
&+{1\over 2}(-1)^pe\bigg[
\hbox{$1\over 2$}h^\kappa{}_\kappa h^\lambda{}_\lambda 
-h^\kappa{}_\lambda h^\lambda{}_\kappa \bigg]
\bar\nabla^\alpha\bar\nabla_\alpha b
-(-1)^pe h^\kappa{}_\kappa\pi\bar\nabla^\alpha\bar\nabla_\alpha b
\vphantom{\bigg\{}&\cr
&+(-1)^p{5D-14+2(D-5)p-2p^2\over 2(2+p)(D-4-p)}e
\pi\pi\bar\nabla^\alpha\bar\nabla_\alpha b
\vphantom{\bigg\{}&\cr
&+h^{\kappa\lambda}\bar\nabla_\kappa\bar\nabla^\alpha b
\bar\nabla_\lambda\bar\nabla_\alpha b
-\hbox{$1\over 2$}h^\kappa{}_\kappa\bar\nabla^\lambda\bar\nabla^\alpha b
\bar\nabla_\lambda\bar\nabla_\alpha b
+{1+p\over 2+p}\pi\bar\nabla^\kappa\bar\nabla^\alpha b
\bar\nabla_\kappa\bar\nabla_\alpha b
\vphantom{\bigg\{}&\cr
&-\hbox{$1\over 2$}h^\kappa{}_\kappa\bar\nabla^\alpha\bar\nabla_\alpha b
\bar\nabla^\beta\bar\nabla_\beta b
+{D-3-p\over D-4-p}\pi\bar\nabla^\alpha\bar\nabla_\alpha b
\bar\nabla^\beta\bar\nabla_\beta b
+\ldots\bigg\}.&(2.8)\cr}
$$
Above, we show the terms containing the fields $h_{\kappa\lambda}$, 
$\pi$, $b$ only, since they contain the relevant scalar fluctuations 
$s$ and $t$ which couple to the chiral operator of the boundary conformal 
field theory of our interest. 

Instead of fixing the gauge right away, we isolate the relevant scalar 
degrees of freedom by performing suitable field redefinitions, as done 
in [17,18]. We write the AdS metric fluctuations as
$$
h_{\kappa\lambda}=\phi_{\kappa\lambda}
+\big(\bar\nabla_\kappa\bar\nabla_\lambda
-\hbox{$1\over D-4-p$}\bar g_{\kappa\lambda}
\bar\nabla^\alpha\bar\nabla_\alpha\big)\varphi,
\eqno(2.9a)
$$
where
$$
\eqalignno{
\varphi&=
\bigg(-\bar\nabla^\alpha\bar\nabla_\alpha
+{(1+p)(D-4-p)\over(D-2)(D-3-p)}e^2\bigg)^{-1}&\cr
&\times\bigg({D-2\over(2+p)(D-3-p)}\pi+(-1)^p{2(D-4-p)\over D-3-p}eb\bigg).
&(2.9b)\cr}
$$
By construction, the field $\phi_{\kappa\lambda}$ decouples from 
the fields $\pi$, $b$ at the quadratic level. 

The scalar fields $s$, $t$ are given by linear functionals of $\pi$, $b$
non local in ${\rm S}_{2+p}$ defined as follows. One expands $s$, $t$ as
well as the scalar fields $\pi$, $b$ with respect to an
orthonormal basis $\{Y_I\}$ of scalar spherical harmonics of ${\rm S}_{2+p}$ 
(cfr. appendix A1) 
$$
\pi=\sum_I\pi_IY_I,\quad b=\sum_I b_IY_I.
\eqno(2.10a)-(2.10b)
$$
Then,
$$
s=\sum_Is_IY_I,\quad t=\sum_I t_IY_I,
\eqno(2.11a)-(2.11b)
$$
where
$$
\eqalignno{
s_I&={1\over 2k+1+p}\bigg({1\over 2(2+p)(D-3-p)}\pi_I
+{(-1)^p(k+1+p)\over (1+p)(D-2)}eb_I\bigg),&(2.12a)\cr
t_I&={1\over 2k+1+p}\bigg({1\over 2(2+p)(D-3-p)}\pi_I
-{(-1)^p k\over (1+p)(D-2)}eb_I\bigg).&(2.12b)\cr}
$$

The action obtained in this way is of the general form
$$
\eqalignno{
I&=\int_{{\rm AdS}_{d+1}}dy^{d+1}(-\bar g_{d+1})^{1\over 2}
\bigg\{-\sum_i{A_i\over 2}\big[\bar\nabla^\kappa\psi_i\bar\nabla_\kappa\psi_i
\vphantom{\bigg\{}+m_i{}^2\psi_i\psi_i\big]&\cr
&+\sum_{ijk}\Big[\lambda_{ijk}\psi_i
\bar\nabla^\kappa\bar\nabla_\kappa\psi_j
\bar\nabla^\lambda\bar\nabla_\lambda\psi_k 
+\mu_{ijk}\psi_i\psi_j
\bar\nabla^\kappa\bar\nabla_\kappa
\bar\nabla^\lambda\bar\nabla_\lambda\psi_k
\vphantom{\bigg\{}&\cr
&+\rho_{ijk}\psi_i\psi_j
\bar\nabla^\kappa\bar\nabla_\kappa\psi_k 
+\sigma_{ijk}\psi_i\psi_j\psi_k\Big]+\cdots\bigg\},&\cr}
$$
where the $\psi_i$ are scalar fields and 
$\lambda_{ijk}=\lambda_{ikj}$, $\mu_{ijk}=\mu_{jik}$, $\rho_{ijk}=\rho_{jik}$,
$\sigma_{ijk}=\sigma_{jik}=\sigma_{ikj}=\cdots$ etc. and the ellipses
denote terms in the $\psi_i$ of order larger than 3.
By performing successively the field redefinitions
$$
\eqalignno{
\psi_i&=\psi_i'-{1 \over A_i}\sum_{jk}\Big[\lambda_{jki}
\psi_j'\bar\nabla^\kappa\bar\nabla_\kappa\psi_k'
+\mu_{jki}\bar\nabla^\kappa\bar\nabla_\kappa(\psi_j'\psi_k')\Big],&\cr
\psi_i'&=\psi_i''-\sum_{jk}\Big(\rho_{jki}+\lambda_{jik}m_k{}^2
+\mu_{jki}m_i{}^2\Big)\psi_j''\psi_k'',&\cr}
$$
one can bring the action in the form
$$
\eqalignno{
I&=\int_{{\rm AdS}_{d+1}}dy^{d+1}(-\bar g_{d+1})^{1\over 2}
\bigg\{-\sum_i
{A_i\over 2}\big[\bar\nabla^\kappa\psi_i''\bar\nabla_\kappa\psi_i''
\vphantom{\bigg\{}+m_i{}^2\psi_i''\psi_i''\big]&\cr
&+\sum_{ijk}g_{ijk}\psi_i''\psi_j''\psi_k''+\cdots\bigg\},&\cr}
$$
where the totally symmetric coupling constants $g_{ijk}$ are given by
$$
\eqalignno{
g_{ijk}&=
\hbox{$1\over 3$}(\lambda_{ijk}m_j{}^2m_k{}^2+\lambda_{jki}m_k{}^2m_i{}^2+
\lambda_{kij}m_i{}^2m_j{}^2)&\cr
&\hphantom{=}
+\hbox{$1\over 3$}(\mu_{ijk}m_k{}^4+\mu_{jki}m_i{}^4+\mu_{kij}m_j{}^4)&\cr
&\hphantom{=}
+\hbox{$1\over 3$}(\rho_{ijk}m_k{}^2+\rho_{jki}m_i{}^2+\rho_{kij}m_j{}^2)
+\sigma_{ijk}.&\cr}
$$

Then, one finds that, after performing the indicated field redefinitions, 
the action of $s$ and $t$ to cubic order is given by
(suppressing double primes for simplicity)
$$
\eqalignno{
&I^{st}_{[\leq 3]}={1\over 4\kappa^2}
\int_{{\rm AdS}_{d+1}}d^{d+1}y (-\bar g_{d+1})^{1\over 2}&\cr
&\bigg\{\sum_I\bigg[
A^s_I\Big(-\hbox{$1\over 2$}\bar\nabla^\kappa s_I\bar\nabla_\kappa s_I
-\hbox{$1\over 2$}m_{sI}{}^2s_Is_I\Big)
+A^t_I\Big(-\hbox{$1\over 2$}\bar\nabla^\kappa t_I\bar\nabla_\kappa t_I
-\hbox{$1\over 2$}m_{tI}{}^2t_It_I\Big)\bigg]
\vphantom{\bigg\{}&\cr
&+\sum_{I_1I_2I_3}\bigg[
g^{sss}_{I_1I_2I_3}s_{I_1}s_{I_2}s_{I_3}
+g^{sst}_{I_1I_2I_3}s_{I_1}s_{I_2}t_{I_3}
+g^{tts}_{I_1I_2I_3}t_{I_1}t_{I_2}s_{I_3}
+g^{ttt}_{I_1I_2I_3}t_{I_1}t_{I_2}t_{I_3}\bigg]\bigg\},~~~~~ &(2.13)\cr}
$$
with $d=D-3-p$. The various constants appearing in the actions 
are given by the following expressions.
$$
\eqalignno{
A^s_I&={2\nu k(k-1)(2k+1+p)\over k+\gamma_s} z_I \bar e^{-2-p},
&(2.14a)\cr
A^t_I&={2\nu (k+1+p)(k+2+p)(2k+1+p)\over k+\gamma_t} z_I \bar e^{-2-p};
&(2.14b)\cr}
$$
$$
\eqalignno{
m_{sI}{}^2&=k(k-1-p)\bar e^2, &(2.15a)\cr
m_{tI}{}^2&=(k+1+p)(k+2+2p)\bar e^2; &(2.15b)\cr}
$$
$$
\eqalignno{
&g^{sss}_{I_1I_2I_3}=\zeta\,
{\alpha_1\alpha_2\alpha_3\big(\alpha - {1\over 2}(1+p)\big) 
\big(\alpha + {1\over 2}(1+p)\big)\over 3(k_1+\gamma_s)(k_2+\gamma_s)(k_3+\gamma_s)} &\cr
&~~\times\bigg\{
(\alpha-1) \Big(\alpha + {1+p\over D-3-p}\Big) 
\Big(\alpha + {(1+p)(-D+4+2p)\over 2(D-2)}\Big)&\cr
&\hphantom{~~\times\bigg\{}
+{\theta\over \nu}\Big[(1+p)(D-3-p)
(\alpha_1{}^2 \alpha_2 + \alpha_2{}^2 \alpha_1 + \alpha_2{}^2 \alpha_3 
+ \alpha_3{}^2 \alpha_2 + \alpha_3{}^2 \alpha_1 + \alpha_1{}^2 \alpha_3)
\vphantom{2\over 3}&\cr
&\hphantom{~~\times\bigg\{}
+ (3D-8 + (2D-8)p - 2p^2)\alpha_1 \alpha_2 \alpha_3 
\vphantom{2\over 3}&\cr
&\hphantom{~~\times\bigg\{}
+ (1+p)(-\hbox{$1\over 2$}D+2+p)(\alpha_1\alpha_2 +\alpha_2 \alpha_3 
+ \alpha_3 \alpha_1)\Big]\bigg\}
a_{I_1 I_2 I_3}\langle{\cal C}_{I_1}{\cal C}_{I_2}{\cal C}_{I_3}\rangle
\bar e^{-p},&(2.16a)\cr
&g^{sst}_{I_1I_2I_3}=
\zeta
\,{\big(\alpha_1 +{1\over 2}(1+p)\big)\big(\alpha_2 +{1\over 2}(1+p)\big)
\alpha_3 (\alpha_3-1-p)\big(\alpha+{1\over 2}(1+p)\big)
\over (k_1+\gamma_s)(k_2+\gamma_s)(k_3+\gamma_t)}&\cr
&~~\times\bigg\{ (\alpha_3-1)
\Big(\alpha_3 + {(1+p)(-D+3+p)\over D-2}\Big)
\Big(\alpha_3 + {(1+p)(-D+4+p)\over D-3-p}\Big)&\cr
&\hphantom{~~\times\bigg\{}
- {\theta\over \nu}\Big[(1+p)(D-3-p)  
(\alpha_1{}^2 \alpha_3 + \alpha_3{}^2 \alpha_1 
+ \alpha_2{}^2 \alpha_3 + \alpha_3{}^2 \alpha_2)
\vphantom{2\over 3}&\cr
&\hphantom{~~\times\bigg\{}  
+ (-1 + (D-4) p - p^2) (\alpha_1{}^2 \alpha_2 +\alpha_2{}^2 \alpha_1) 
\vphantom{2\over 3}&\cr
&\hphantom{~~\times\bigg\{}  
+ (D-4 + (2D-8) p - 2 p^2) \alpha_1 \alpha_2 \alpha_3 
\vphantom{2\over 3}&\cr
&\hphantom{~~\times\bigg\{}
+ (1+p) p (D-3-p)(\alpha_1 \alpha_3 + \alpha_2 \alpha_3) 
\vphantom{2\over 3}&\cr
&\hphantom{~~\times\bigg\{}
+ (1+p) (-D+2 + (D-3) p - p^2) \alpha_1 \alpha_2 
\vphantom{2\over 3}&\cr
&\hphantom{~~\times\bigg\{} 
+(1+p) (D-3-p) (-\alpha_1{}^2 - \alpha_2{}^2 + (1+p)\alpha_3{}^2)
\vphantom{2\over 3}&\cr
&\hphantom{~~\times\bigg\{}
+ (1+p)^2 (-D+3+p) (\alpha_1 + \alpha_2 +\alpha_3)
\Big]\bigg\}
a_{I_1 I_2 I_3}\langle{\cal C}_{I_1}{\cal C}_{I_2}{\cal C}_{I_3}\rangle
\bar e^{-p},&(2.16b)\cr
&g^{tts}_{I_1I_2I_3}=
\zeta 
\,{\alpha_1 \alpha_2 \big(\alpha_3 +{1\over 2}(1+p)\big)  
\big(\alpha_3 + {3\over2}(1+p)\big)(\alpha+1+p)
\over (k_1+\gamma_t)(k_2+\gamma_t)(k_3+\gamma_s)}&\cr
&~~\times\bigg\{ (\alpha_3+2+p)
\Big(\alpha_3 + {(1+p)(3D-8-2p)\over 2 (D-2)}\Big)
\Big(\alpha_3 + {(1+p)(D-4-p)\over D-3-p}\Big)&\cr
&\hphantom{~~\times\bigg\{}
- {\theta\over \nu}\Big[(1+p)(D-3-p)  
(\alpha_1{}^2 \alpha_3 + \alpha_3{}^2 \alpha_1 
+ \alpha_2{}^2 \alpha_3 + \alpha_3{}^2 \alpha_2)
\vphantom{2\over 3}&\cr
&\hphantom{~~\times\bigg\{}  
+  (-1 + (D-4) p - p^2) (\alpha_1{}^2 \alpha_2 +\alpha_2{}^2 \alpha_1) 
\vphantom{2\over 3}&\cr
&\hphantom{~~\times\bigg\{}  
+  (D-4 + (2D-8) p - 2 p^2) \alpha_1 \alpha_2 \alpha_3 
\vphantom{2\over 3}&\cr
&\hphantom{~~\times\bigg\{}
+ (1+p) (\hbox{$5\over 2$}D-8 + (2D- 9) p - 2 p^2)
(\alpha_1 \alpha_3 + \alpha_2 \alpha_3) 
\vphantom{2\over 3}&\cr
&\hphantom{~~\times\bigg\{}
+ (1+p) (\hbox{$3\over 2$}D-6 + (2D-9) p - 2 p^2) \alpha_1 \alpha_2 
\vphantom{2\over 3}&\cr
&\hphantom{~~\times\bigg\{} 
+ (1+p) (\hbox{$3\over 2$}D-5 + (D-5) p -  p^2) (\alpha_1{}^2 + \alpha_2{}^2)
\vphantom{2\over 3}&\cr
&\hphantom{~~\times\bigg\{}
+ (1+p)^2 (\hbox{$3\over 2$}D -5 + (D-5) p -  p^2) (\alpha_1 + \alpha_2)\Big]
\bigg\}
a_{I_1 I_2 I_3}\langle{\cal C}_{I_1}{\cal C}_{I_2}{\cal C}_{I_3}\rangle
\bar e^{-p},&(2.16c)\cr
&g^{ttt}_{I_1I_2I_3}
=\zeta\,{\big(\alpha_1 + {1\over 2}(1+p)\big)
\big(\alpha_2 + {1\over 2}(1+p)\big)\big(\alpha_3+ {1\over 2}(1+p)\big) 
(\alpha + 1 + p) (\alpha +2 + 2p)\over 3(k_1+\gamma_t)(k_2+\gamma_t)(k_3+\gamma_t)} &\cr
&~~\times\bigg\{
(\alpha + 2 +p) \Big(\alpha + {(1+p)(2D-8-2p)\over D-3-p}\Big)
\Big(\alpha + {(1+p)(4D - 9 - p)\over 2(D-2)}\Big)&\cr
&\hphantom{~~\times\bigg\{}
+{\theta\over \nu}\Big[(1+p)(D-3-p)
(\alpha_1{}^2 \alpha_2 + \alpha_2{}^2 \alpha_1 + \alpha_2{}^2 \alpha_3 
+ \alpha_3{}^2 \alpha_2 + \alpha_3{}^2 \alpha_1 + \alpha_1{}^2 \alpha_3)
\vphantom{2\over 3}&\cr
&\hphantom{~~\times\bigg\{}
+ (3D-8 + (2D-8)p - 2p^2)\alpha_1 \alpha_2 \alpha_3 
\vphantom{2\over 3}&\cr
&\hphantom{~~\times\bigg\{}
+ (1+p)(4D-13 + (3D-14)p - 3p^2)(\alpha_1\alpha_2 +\alpha_2 \alpha_3 
+ \alpha_3 \alpha_1)
\vphantom{2\over 3}&\cr
&\hphantom{~~\times\bigg\{}
+ (1+p)^2 (D -\hbox{$7\over 2$} -p)(\alpha_1{}^2 +\alpha_2{}^2 +\alpha_3{}^2)
\vphantom{2\over 3}&\cr
&\hphantom{~~\times\bigg\{}
+ (1+p)^2 (3D -11  +(2D-\hbox{$21\over 2$})p -2 p^2)
(\alpha_1+\alpha_2+\alpha_3)
\vphantom{2\over 3}&\cr
&\hphantom{~~\times\bigg\{}
+(2 +p)(1+p)^3 (D-4-p)\Big]\bigg\}
a_{I_1 I_2 I_3}\langle{\cal C}_{I_1}{\cal C}_{I_2}{\cal C}_{I_3}\rangle
\bar e^{-p},&(2.16d)\cr}
$$
where
$$
\eqalignno{
\nu&=(D-2)(1+p)(D-3-p),\vphantom{2\over 3}&(2.17a)\cr 
\gamma_s&={1+p\over D-3-p},\vphantom{2\over 3}&(2.17b)\cr 
\gamma_t&={(1+p)(D-4-p)\over D-3-p}, \vphantom{2\over 3}&(2.17c)\cr
\theta&=-D+1+(D-4)p-p^2,\vphantom{2\over 3}&(2.17d)\cr
\zeta&= 16(D-2)^2(1+p)(D-3-p),\vphantom{2\over 3}&(2.17e)\cr}
$$
$$
\bar e^2={(D-3-p)\over (D-2)(1+p)}e^2,
\eqno(2.18)
$$
and 
$$
\eqalignno{
\alpha_1&=\hbox{$1\over 2$}(k_2+k_3-k_1), \quad\hbox{etc.}, &(2.19a)\cr
\alpha&=\hbox{$1\over 2$}(k_1+k_2+k_3),&(2.19b)\cr}
$$
and $z_I$, $a_{I_1 I_2 I_3}$, 
$\langle{\cal C}_{I_1}{\cal C}_{I_2}{\cal C}_{I_3}\rangle$ are defined in 
appendix A1.

As previously mentioned, the fields $q$ and $l_\kappa$
act as Lagrange multipliers enforcing certain 
constraints involving $h_{\kappa\lambda}$, $\pi$, $b$.
Such constraints, however, do not affect the quadratic and cubic terms of 
the fields $s$ and $t$ in the action, which are thus gauge invariant.
This can be shown as follows. 

To linear order, the constraints can be read off from the part 
$I^{\rm constr}_{[2]}$ of the quadratic action $I_{[2]}$ which is linear 
in the fields directly involved by gauge fixing such as 
$l_\kappa$, $n_\alpha$, $q$, etc. By direct computation, one 
can check that only $l_\kappa$, $q$ contribute to $I^{\rm constr}_{[2]}$
$$
\eqalignno{
I^{\rm constr}_{[2]}&={1\over 4\kappa^2}
\int_{{\rm AdS}_{D-2-p}}d^{D-2-p}y (-\bar g_{D-2-p})^{1\over 2}
\int_{{\rm S}_{2+p}}d^{2+p}x (\bar g_{2+p})^{1\over 2}&\cr
&\bigg\{{1\over 2}\bigg[{1+p\over 2+p}
\bar\nabla^\alpha\bar\nabla_\alpha +{D-3-p\over D-2}e^2\bigg]
\bar\nabla^\beta\bar\nabla_\beta q
\bigg[h^\kappa{}_\kappa-{2(D-2)\over (2+p)(D-4-p)}\pi\bigg]
\vphantom{\bigg\{}&\cr
&+\bar\nabla^\alpha\bar\nabla_\alpha l^\kappa
\bigg[{(D-2)\over (2+p)(D-4-p)}\bar\nabla_\kappa\pi
+2(-1)^pe\bar\nabla_\kappa b +
\bar\nabla^\lambda h_{\lambda\kappa}-\bar\nabla_\kappa h^\lambda{}_\lambda
\bigg]\bigg\}.\vphantom{\bigg\{}~~~~~~&(2.20)\cr}
$$
Perform the field redefinition (2.9) and expand $\phi_{\kappa\lambda}$
$l_\kappa$, $q$ in scalar harmonics of ${\rm S}_{2+p}$
$$
\phi_{\kappa\lambda}=\sum_I\phi_{I\kappa\lambda}Y_I,
\eqno(2.21)
$$
$$
q=\sum_I q_IY_I,\quad
l_\kappa=\sum_I l_{I\kappa}Y_I.
\eqno(2.22a)-(2.22b)
$$
Then $I^{\rm constr}_{[2]}$ takes the form
$$
\eqalignno{
I^{\rm constr}_{[2]}&={1\over 4\kappa^2}
\int_{{\rm AdS}_{d+1}}d^{d+1}y (-\bar g_{d+1})^{1\over 2}&\cr
&\sum_I\bigg\{
u_Iq_I\Big[\phi_I{}^\kappa{}_\kappa
+v^s_I\Big(\bar\nabla^\kappa\bar\nabla_\kappa s_I-m_{sI}{}^2s_I\Big)
+v^t_I\Big(\bar\nabla^\kappa\bar\nabla_\kappa t_I-m_{tI}{}^2t_I\Big)\Big]&\cr
&-w_Il_I{}^\kappa\Big[\bar\nabla^\lambda\phi_{I\lambda\kappa}
-\bar\nabla_\kappa\phi_I{}^\lambda{}_\lambda\Big]\bigg\},
&(2.23)\cr}
$$
where
$$
\eqalignno{
u_I&={1+p\over 2(2+p)}(k-1)k(k+1+p)(k+2+p)z_I \bar e^{2-p},&(2.24)\cr
v^s_{I}&={2(D-2)\over k+\gamma_s}\bar e^{-2},&(2.25a)\cr
v^t_{I}&={2(D-2)\over k+\gamma_t}\bar e^{-2},&(2.25b)\cr
w_I&=k(k+1+p)z_I \bar e^{-p},&(2.26)\cr}
$$
$\gamma_s$, $\gamma_t$ are given by (2.17b), (2.17c)
and $z_I$ is defined in appendix A1.
The gauge fixing conditions (2.5) would amount to
$$
l_{I\kappa}=0, \quad k\geq 1,\quad\quad q_I=0,\quad k\geq 2.
\eqno(2.27)
$$
The constraint associated to $q_I$ yields an apparent mixing of 
$\phi_I{}^\kappa{}_\kappa$, $s_I$, $t_I$. The mixing can be 
removed, before fixing the gauge, 
by performing the following field redefinition
in the action $I^{st}_{[\leq 3]}$ and 
$I^{\rm constr}_{[2]}$ (cfr eqs. (2.13) and (2.23)):
$$
s_I=s'_I-c^s_Iq_I,\quad t_I=t'_I-c^t_Iq_I,
\eqno(2.28)
$$
where
$$
\eqalignno{
c^s_I&={(k+1+p)(k+2+p)\over 2(D-3-p)(2+p)(2k+1+p)}\bar e^2,&(2.29a)\cr
c^t_I&={k(k-1)\over 2(D-3-p)(2+p)(2k+1+p)}\bar e^2.&(2.29b)\cr}
$$
Then, the $s$ $t$ action $I^{st}_{[\leq 3]}$ remains of the form (2.13)
while $I^{\rm constr}_{[2]}$ becomes
$$
I^{\rm constr}_{[2]}={1\over 4\kappa^2}
\int_{{\rm AdS}_{d+1}}d^{d+1}y (-\bar g_{d+1})^{1\over 2}
\sum_I\bigg\{u_Iq_I\phi_I{}^\kappa{}_\kappa
-w_Il_I{}^\kappa\Big[\bar\nabla^\lambda\phi_{I\lambda\kappa}
-\bar\nabla_\kappa\phi_I{}^\lambda{}_\lambda\Big]\bigg\}.
\eqno(2.30)
$$
In this way, the linear order constraints are simply conditions
on $\phi_{\kappa\lambda}$ and decouple from $s$ and $t$.
At higher orders,
the constraints yielded by gauge fixing are therefore of the form
$$
\eqalignno{
\phi^\kappa{}_\kappa&=C(\phi,s,t)+\ldots, &(2.31a)\cr
&&\cr
\bar\nabla^\lambda\phi_{\lambda\kappa}
-\bar\nabla_\kappa\phi^\lambda{}_\lambda&=D(\phi,s,t)+\ldots, &(2.31b)\cr}
$$
where $C(\phi,s,t)$, $D(\phi,s,t)$ are certain composites of the fields 
$\phi_{\kappa\lambda}$, $s$, $t$ of polynomial degree at least 2
and the ellipses denote contributions containing other fields. 
Thus, it is evident that the $\phi\phi$ terms of the quadratic action yield 
no $sss$, $sst$, $tts$, $ttt$ couplings at cubic level via the constraints.
Since there are no $\phi\pi$, $\phi b$ terms in the quadratic action, 
the constraints cannot generate any $sss$, $sst$, $tts$, $ttt$ couplings
from these at cubic level. Higher order terms involving $\phi$ produce
obviously no contribution to the cubic coupling terms of $s$ and $t$.
Thus, the $s$, $t$ action $I^{st}_{[\leq 3]}$ is not affected by the 
constraints associated with gauge fixing as announced.

From (2.13) and (2.17a), it appears that the kinetic term of the fields $s_I$ 
with $k=0,1$ vanishes. Therefore, such fields are non propagating
gauge degrees of freedom. Consistency requires that they should decouple from 
all the propagating physical fields. Indeed it is easy to check that 
the coupling constants $g^{sss}_{I_1I_2I_3}$, $g^{sst}_{I_1I_2I_3}$, 
$g^{tts}_{I_1I_2I_3}$ (cfr. eqs. (2.16a)--(2.16c)) all vanish whenever 
the fields $s_I$ with $k=0,1$ are involved. To this end, one has 
to use the property that the triple 
contraction $\langle{\cal C}_{I_1}{\cal C}_{I_2}{\cal C}_{I_3}\rangle$
vanishes unless the values of the non negative integers $k_1$, $k_2$, $k_3$ 
are such to allow complete contraction of the $SO(3+p)$ indices
of the ${\cal C}_{I}$'s (see appendix A1).

The expressions of the coupling constants $g_{I_1I_2I_3}$ simplify
considerably when the constant $\theta$ (cfr. eq. (2.17d)) vanishes.
This happens precisely for $(D,p)=(11,2)$, $(11,5)$, $(10,3)$. These 
values correspond to the physically interesting cases
of ${\rm AdS}_7\times{\rm S}_4$, ${\rm AdS}_4\times{\rm S}_7$,
${\rm AdS}_5\times{\rm S}_5$.

Writing in eq. (2.13)
$$
A_I=\bar A_I z_I \bar e^{-2-p},
\eqno(2.32)
$$
$$
m_I{}^2=\bar m_I{}^2 \bar e^2,
\eqno(2.33)
$$
$$
g_{I_1I_2I_3}=\bar g_{I_1I_2I_3}
a_{I_1 I_2 I_3}\langle{\cal C}_{I_1}{\cal C}_{I_2}{\cal C}_{I_3}\rangle
\bar e^{-p},
\eqno(2.34)
$$
one has:
\par\vskip.6cm

$\underline{{\rm AdS}_7\times{\rm S}_4}$

$$
\eqalignno{
\bar A^s_I&={324 k(k-1)(2k+3)\over k+{1\over 2}},
&(2.35a)\cr
\bar A^t_I&={324 (k+3)(k+4)(2k+3)\over k+{5\over 2}};
&(2.35b)\cr}
$$
$$
\eqalignno{
\bar m_{sI}^2&=k(k-3),&(2.36a)\cr
\bar m_{tI}^2&=(k+3)(k+6);&(2.36b)\cr}
$$
$$\eqalignno{
\bar g^{sss}_{I_1I_2I_3}&=
{7776\alpha_1\alpha_2\alpha_3
(\alpha-1)(\alpha^2-{1\over4})(\alpha^2-{9\over 4})
\over (k_1+{1\over 2})(k_2+{1\over 2})(k_3+{1\over 2})},
&(2.37a)\cr
\bar g^{sst}_{I_1I_2I_3}&=
{23328(\alpha_1+{3\over 2})(\alpha_2+{3\over 2})\alpha_3 
(\alpha_3-1)(\alpha_3-2)(\alpha_3-{5\over2})(\alpha_3-3)(\alpha+{3\over2})
\over (k_1+{1\over 2})(k_2+{1\over 2})(k_3+{5\over 2})},
&(2.37b)\cr
\bar g^{tts}_{I_1I_2I_3}&=
{23328\alpha_1\alpha_2(\alpha_3+{3\over 2})(\alpha_3+{5\over 2})
(\alpha_3+{7\over 2})(\alpha_3+4)(\alpha_3+{9\over 2})(\alpha+3)
\over (k_1+{5\over 2})(k_2+{5\over 2})(k_3+{1\over 2})},
&(2.37c)\cr
\bar g^{ttt}_{I_1I_2I_3}&=
{7776(\alpha_1+{3\over 2})(\alpha_2+{3\over 2})(\alpha_3+{3\over 2}) 
(\alpha+3)(\alpha+4)(\alpha+5)(\alpha+{11\over 2})(\alpha+6)
\over (k_1+{5\over 2})(k_2+{5\over 2})(k_3+{5\over 2})};~~~~~~
&(2.37d)\cr}
$$

$\underline{{\rm AdS}_4\times{\rm S}_7}$

$$
\eqalignno{
\bar A^s_I&={324 k(k-1)(2k+6)\over k+2},
&(2.38a)\cr
\bar A^t_I&={324 (k+6)(k+7)(2k+6)\over k+4};
&(2.38b)\cr}
$$
$$
\eqalignno{
\bar m_{sI}^2&=k(k-6),&(2.39a)\cr
\bar m_{tI}^2&=(k+6)(k+12);&(2.39b)\cr}
$$
$$\eqalignno{
\bar g^{sss}_{I_1I_2I_3}&=
{7776\alpha_1\alpha_2\alpha_3(\alpha+2)(\alpha^2-1)(\alpha^2-9)
\over (k_1+2)(k_2+2)(k_3+2)},
&(2.40a)\cr
\bar g^{sst}_{I_1I_2I_3}&=
{23328(\alpha_1+3)(\alpha_2+3)\alpha_3 
(\alpha_3-1)(\alpha_3-2)(\alpha_3-4)(\alpha_3-6)(\alpha+3)
\over (k_1+2)(k_2+2)(k_3+4)},
&(2.40b)\cr
\bar g^{tts}_{I_1I_2I_3}&=
{23328\alpha_1\alpha_2(\alpha_3+3)
(\alpha_3+4)(\alpha_3+5)(\alpha_3+7)(\alpha_3+9)(\alpha+6)
\over (k_1+4)(k_2+4)(k_3+2)},
&(2.40c)\cr
\bar g^{ttt}_{I_1I_2I_3}&=
{7776(\alpha_1+3)(\alpha_2+3)(\alpha_3+3) 
(\alpha+6)(\alpha+7)(\alpha+8)(\alpha+10)(\alpha+12)
\over (k_1+4)(k_2+4)(k_3+4)};~~~~~~
&(2.40d)\cr}
$$

$\underline{{\rm AdS}_5\times{\rm S}_5}$

$$
\eqalignno{
\bar A^s_I&={256k(k-1)(2k+4)\over k+1},
&(2.41a)\cr
\bar A^t_I&={256(k+4)(k+5)(2k+4)\over k+3};
&(2.41b)\cr}
$$
$$
\eqalignno{
\bar m_{sI}^2&=k(k-4),&(2.42a)\cr
\bar m_{tI}^2&=(k+4)(k+8);&(2.42b)\cr}
$$
$$\eqalignno{
\bar g^{sss}_{I_1I_2I_3}&=
{16384\alpha_1\alpha_2\alpha_3\alpha(\alpha^2-1)(\alpha^2-4)
\over 3(k_1+1)(k_2+1)(k_3+1)},
&(2.43a)\cr
\bar g^{sst}_{I_1I_2I_3}&=
{16384(\alpha_1+2)(\alpha_2+2)\alpha_3 
(\alpha_3-1)(\alpha_3-2)(\alpha_3-3)(\alpha_3-4)(\alpha+2)
\over (k_1+1)(k_2+1)(k_3+3)},
&(2.43b)\cr
\bar g^{tts}_{I_1I_2I_3}&=
{16384\alpha_1\alpha_2(\alpha_3+2)
(\alpha_3+3)(\alpha_3+4)(\alpha_3+5)(\alpha_3+6)(\alpha+4)
\over (k_1+3)(k_2+3)(k_3+1)},
&(2.43c)\cr
\bar g^{ttt}_{I_1I_2I_3}&=
{16384(\alpha_1+2)(\alpha_2+2)(\alpha_3+2) 
(\alpha+4)(\alpha+5)(\alpha+6)(\alpha+7)(\alpha+8)
\over 3(k_1+3)(k_2+3)(k_3+3)}.~~~~~~
&(2.43d)\cr}
$$

\par\vskip.6cm
\item{\bf 3.} {\bf Application of the method}
\vskip.4cm
\par
We are now ready to compute two and three point functions
in the SCFTs using the ${\rm AdS}_{d+1}/{\rm CFT}_d$ correspondence.
The general formulas derived in [19] work with AdS radius set to 1.
Assume that the AdS scalar fields $\phi_i$ correspond to the CFT 
local field ${\cal O}_i$. The mass $m_i$ of $\phi_i$ and the 
conformal dimension $\Delta_i$ of ${\cal O}_i$ are related as
$$
\Delta_i=\hbox{$1\over2$}\Big[d+\Big(d^2+4m_i{}^2\Big)^{1\over 2}\Big].
\eqno(3.1)
$$
Then
$$
\langle {\cal O}_i(x) {\cal O}_j(y)\rangle = 
{2\over \pi^{d\over2}} \eta_i{\Delta_i - {d\over 2} \over \Delta_i} 
{\Gamma(\Delta_i +1)\over \Gamma(\Delta_i -{d\over2})}
{(w_i)^2 \delta_{ij} \over |x-y|^{2\Delta_i}},
\eqno(3.2)
$$
where $\eta_i$ is the coefficient of the canonically normalized 
kinetic term of the bulk field $\phi_i$,
and 
$$
\langle {\cal O}_i(x) {\cal O}_j(y) {\cal O}_k(z) \rangle = 
{R_{ijk}\over |x-y|^{\Delta_i+\Delta_j-\Delta_k}
|y-z|^{\Delta_j+\Delta_k-\Delta_i}
|z-x|^{\Delta_k+\Delta_i-\Delta_j}},
\eqno(3.3)
$$
with
$$
\eqalignno{
R_{ijk}=&{1\over 2\pi^d}\lambda_{ijk}
{\Gamma({1\over2}(\Delta_i+\Delta_j-\Delta_k))
\Gamma({1\over2}(\Delta_j+\Delta_k-\Delta_i))
\Gamma({1\over2}(\Delta_k+\Delta_i-\Delta_j))
\over 
\Gamma(\Delta_i -{d\over2})
\Gamma(\Delta_j -{d\over2})
\Gamma(\Delta_k -{d\over2})} 
&\cr 
&\Gamma(\hbox{${1\over2}$}({\Delta_i+\Delta_j+\Delta_k} - d))
w_iw_jw_k,&(3.4)\cr}
$$
where $\lambda_{ijk}$ is the cubic coupling constant of 
$\phi_i$, $\phi_j$, $\phi_k$
multiplied by the appropriate symmetry factor. 
The factors $w_i$ parametrize unknown proportionality constants
which relate the fields $\phi_i$ to the sources of the operators
${\cal O}_i$, as in [5].
These factors can presumably be fixed by carefully 
studying absorption processes on the branes [20]. 
However, for the present purposes we follow ref. [5]
and fix them to normalize the two point functions as
$$
\langle {\cal O}_i(x) {\cal O}_j(y) \rangle = 
{\delta_{ij} \over |x-y|^{2\Delta_i}}.
\eqno(3.5)
$$
With this canonical normalization the three point functions are readily 
computed.

In our case, $d=D-3-p$.
Imposing that the AdS radius is 1 fixes the value of $\bar e$ to be
$$
\bar e={d\over 1+p}.
\eqno(3.6)
$$
We denote by ${\cal O}^s_I$, ${\cal O}^t_I$ the ${\rm CFT}_d$
operators corresponding to the ${\rm AdS}_{d+1}$ scalars $s_I$, $t_I$ in 
the ${\rm AdS}_{d+1}/{\rm CFT}_d$  duality. Their dimensions are given by
$$
\eqalignno{
\Delta^s_I&={dk\over 1+p},&(3.7a)\cr
\Delta^t_I&={d(k+2+2p)\over 1+p}.&(3.7b)\cr}
$$
From (2.13), it appears that in the present case
$$
\eqalignno{
\eta^s_I&={1\over 4\kappa^2}A^s_{I_1},\quad \hbox{etc.},&(3.8)\cr
\lambda^{sss}_{I_1I_2I_3}&={3!\over 4\kappa^2} g^{sss}_{I_1I_2I_3},\quad
\lambda^{sst}_{I_1I_2I_3}={2!\over 4\kappa^2} g^{sst}_{I_1I_2I_3},\quad 
\hbox{etc.}.&(3.9)\cr}
$$

We find the following expressions.

\vskip.2cm

$\underline{{\rm AdS}_7\times{\rm S}_4}$

\vskip.2cm

In this case one has ${1\over 4\kappa^2} = {2 N^3 \over \pi^5}$ and $\bar e=2$.
Set 
$$
\eqalignno{
\sigma(k)&=\big[(2k-2)!\big]^{-{1\over 2}},&(3.10a)\cr
\tau(k)&=16\bigg[{(2k+1)!(2k+4)!(2k+7)!\over 
(2k+6)!(2k+8)!(2k+9)!(2k+11)!}\bigg]^{{1\over 2}}.
&(3.10b)\cr}
$$
Then,
$$
\eqalignno{
R^{sss}_{I_1I_2I_3}&={1\over 4(\pi N)^{3\over 2}}
\langle{\cal C}_{I_1}{\cal C}_{I_2}{\cal C}_{I_3}\rangle 
\sigma(k_1)\sigma(k_2)\sigma(k_3)
2^{2\alpha}\Gamma(\alpha)&\cr
&\times\vphantom{{1\over 4(\pi N)^{3\over 2}}} 
\Gamma(\alpha_1+\hbox{$1\over 2$})
\Gamma(\alpha_2+\hbox{$1\over 2$})\Gamma(\alpha_3+ \hbox{$1\over 2$}),
&(3.11a)\cr
R^{sst}_{I_1I_2I_3}&={1\over 4(\pi N)^{3\over 2}}
\langle{\cal C}_{I_1}{\cal C}_{I_2}{\cal C}_{I_3}\rangle
\sigma(k_1)\sigma(k_2)\tau(k_3)
2^{2\alpha}\Gamma(\alpha+2)&\cr
&\times
{\Gamma(2\alpha_1+6)\Gamma(2\alpha_2+6)
\over \Gamma(2\alpha_1+2)\Gamma(2\alpha_2+2)}
\Gamma(\alpha_1+\hbox{$5\over 2$})\Gamma(\alpha_2+\hbox{$5\over 2$})
\Gamma(\alpha_3-\hbox{$3\over 2$}),
&(3.11b)\cr
R^{tts}_{I_1I_2I_3}&={1\over 4(\pi N)^{3\over 2}}
\langle{\cal C}_{I_1}{\cal C}_{I_2}{\cal C}_{I_3}\rangle 
\tau(k_1)\tau(k_2)\sigma(k_3)
2^{2\alpha}{\Gamma(2\alpha+9)\over\Gamma(2\alpha+5)}\Gamma(\alpha+4)&\cr
&\times
{\Gamma(2\alpha_3+10)\Gamma(2\alpha_3+12)
\over \Gamma(2\alpha_3+2)\Gamma(2\alpha_3+8)}
\Gamma(\alpha_1+\hbox{$1\over 2$})\Gamma(\alpha_2+\hbox{$1\over 2$})
\Gamma(\alpha_3+\hbox{$9\over 2$}),
&(3.11c)\cr
R^{ttt}_{I_1I_2I_3}&={1\over 4(\pi N)^{3\over 2}}
\langle{\cal C}_{I_1}{\cal C}_{I_2}{\cal C}_{I_3}\rangle     
\tau(k_1)\tau(k_2)\tau(k_3)
2^{2\alpha}{\Gamma(2\alpha+13)\Gamma(2\alpha+15)
\over \Gamma(2\alpha+5)\Gamma(2\alpha+11)}
\Gamma(\alpha+6)&\cr
&\times
{\Gamma(2\alpha_1+6)\Gamma(2\alpha_2+6)\Gamma(2\alpha_3+6)
\over \Gamma(2\alpha_1+2)\Gamma(2\alpha_2+2)\Gamma(2\alpha_3+2)}
\Gamma(\alpha_1+\hbox{$5\over 2$})\Gamma(\alpha_2+\hbox{$5\over 2$})
\Gamma(\alpha_3+\hbox{$5\over 2$}).
&(3.11d)\cr}
$$

\vskip.2cm

$\underline{{\rm AdS}_4\times{\rm S}_7}$

\vskip.2cm

In this case one has ${1\over 4\kappa^2} = {N^{3\over 2} 
\over 2^{19\over 2} \pi^5}$ and $\bar e={1\over 2}$. Set 
$$
\eqalignno{
\sigma(k)&=\big[(k+1)!\big]^{1\over 2},&(3.12a)\cr
\tau(k)&={1\over 4}\bigg[{k!(k+2)!(k+3)!(k+5)! \over 
(k+4)!(k+7)!(k+10)!}\bigg]^{{1\over 2}}.&(3.12b)\cr}
$$
Then, 
$$
\eqalignno{
R^{sss}_{I_1I_2I_3}&={\pi\over 2}\Big({2\over N}\Big)^{3\over 4}
\langle{\cal C}_{I_1}{\cal C}_{I_2}{\cal C}_{I_3}\rangle 
\sigma(k_1)\sigma(k_2)\sigma(k_3)
2^{-\alpha}{1\over\Gamma({1\over 2}\alpha+1)}&\cr
&\times{1\over 
\Gamma({1\over 2}(\alpha_1+1))\Gamma({1\over 2}(\alpha_2+1))
\Gamma({1\over 2}(\alpha_3+1))},
&(3.13a)\cr
R^{sst}_{I_1I_2I_3}&={\pi\over 2}\Big({2\over N}\Big)^{3\over 4}
\langle{\cal C}_{I_1}{\cal C}_{I_2}{\cal C}_{I_3}\rangle
\sigma(k_1)\sigma(k_2)\tau(k_3)
2^{-\alpha}{1\over\Gamma({1\over 2}\alpha+2)}&\cr
&\times
{\Gamma(\alpha_1+5)\Gamma(\alpha_2+5)\over 
\Gamma(\alpha_1+1)\Gamma(\alpha_2+1)}
{1\over \Gamma({1\over 2}(\alpha_1+3))\Gamma({1\over 2}(\alpha_2+3))
\Gamma({1\over 2}(\alpha_3-1))},
&(3.13b)\cr
R^{tts}_{I_1I_2I_3}&={\pi\over 2}\Big({2\over N}\Big)^{3\over 4}
\langle{\cal C}_{I_1}{\cal C}_{I_2}{\cal C}_{I_3}\rangle 
\tau(k_1)\tau(k_2)\sigma(k_3)
2^{-\alpha}{\Gamma(\alpha+8)\over\Gamma(\alpha+4)}
{1\over\Gamma({1\over 2}\alpha+3)}&\cr
&\times
{\Gamma(\alpha_3+5)\Gamma(\alpha_3+11)\over 
\Gamma(\alpha_3+1)\Gamma(\alpha_3+3)}
{1\over \Gamma({1\over 2}(\alpha_1+1))
\Gamma({1\over 2}(\alpha_2+1))\Gamma({1\over 2}(\alpha_3+5))},
&(3.13c)\cr
R^{ttt}_{I_1I_2I_3}&={\pi\over 2}\Big({2\over N}\Big)^{3\over 4}
\langle{\cal C}_{I_1}{\cal C}_{I_2}{\cal C}_{I_3}\rangle     
\tau(k_1)\tau(k_2)\tau(k_3)
2^{-\alpha}
{\Gamma(\alpha+8)\Gamma(\alpha+14)\over\Gamma(\alpha+4)\Gamma(\alpha+6)}
{1\over\Gamma({1\over 2}\alpha+4)}&\cr
&\times
{\Gamma(\alpha_1+5)\Gamma(\alpha_2+5)\Gamma(\alpha_3+5)
\over \Gamma(\alpha_1+1)\Gamma(\alpha_2+1)\Gamma(\alpha_3+1)}
{1\over \Gamma({1\over 2}(\alpha_1+3))
\Gamma({1\over 2}(\alpha_2+3))\Gamma({1\over 2}(\alpha_3+3))}.
&(3.13d)\cr}
$$

\vskip.2cm

$\underline{{\rm AdS}_5\times{\rm S}_5}$

\vskip.2cm

In this case one has ${1\over 4\kappa^2} = {N^2 \over 8 \pi^5}$ and 
$\bar e=1$. Set
$$
\eqalignno{
\sigma(k)&=k^{1\over 2},&(3.14a)\cr
\tau(k)&=\bigg[{k!(k+1)!(k+2)!\over 
(k+5)!(k+6)!(k+7)!}\bigg]^{{1\over 2}}.&(3.14b)\cr}
$$
Then, 
$$
\eqalignno{
R^{sss}_{I_1I_2I_3}&={1\over N}
\langle{\cal C}_{I_1}{\cal C}_{I_2}{\cal C}_{I_3}\rangle
\sigma(k_1)\sigma(k_2)\sigma(k_3),
&(3.15a)\cr
R^{sst}_{I_1I_2I_3}&={1\over N}
\langle{\cal C}_{I_1}{\cal C}_{I_2}{\cal C}_{I_3}\rangle
\sigma(k_1)\sigma(k_2)\tau(k_3)&\cr
&\times
{\Gamma(\alpha_1+3)\Gamma(\alpha_1+4) 
\Gamma(\alpha_2+3)\Gamma(\alpha_2+4)\over
\Gamma(\alpha_1+1)\Gamma(\alpha_1+2)
\Gamma(\alpha_2+1)\Gamma(\alpha_2+2)},
&(3.15b)\cr
R^{tts}_{I_1I_2I_3}&={1\over N}
\langle{\cal C}_{I_1}{\cal C}_{I_2}{\cal C}_{I_3}\rangle
\tau(k_1)\tau(k_2)\sigma(k_3)
{\Gamma(\alpha+5)\Gamma(\alpha+6)\over\Gamma(\alpha+3)\Gamma(\alpha+4)}&\cr
&\times
{\Gamma(\alpha_3+7)\Gamma(\alpha_3+8)\over 
\Gamma(\alpha_3+1)\Gamma(\alpha_3+2)},
&(3.15c)\cr
R^{ttt}_{I_1I_2I_3}&={1\over N}
\langle{\cal C}_{I_1}{\cal C}_{I_2}{\cal C}_{I_3}\rangle
\tau(k_1)\tau(k_2)\tau(k_3)
{\Gamma(\alpha+9)\Gamma(\alpha+10)\over\Gamma(\alpha+3)\Gamma(\alpha+4)}&\cr
&\times
{\Gamma(\alpha_1+3)\Gamma(\alpha_1+4)\Gamma(\alpha_2+3)\Gamma(\alpha_2+4)
\Gamma(\alpha_3+3)\Gamma(\alpha_3+4)
\over \Gamma(\alpha_1+1)\Gamma(\alpha_1+2)\Gamma(\alpha_2+1)\Gamma(\alpha_2+2)
\Gamma(\alpha_3+1)\Gamma(\alpha_3+2)}.
&(3.15d)\cr}
$$
\par\vskip.6cm
\item{\bf 4.} {\bf Conclusions}
\vskip.4cm
\par
We have derived three point functions for a set of chiral operators,
including the primary ones, for the large $N$ limit of maximally 
supersymmetric CFT$_d$ in $d=3,4,6$ using the AdS/CFT correspondence. 
In obtaining such results we have used a general gravitational 
action which could treat the different cases simultaneously.
We have obtained the three point couplings for the bulk fields on 
AdS working at the level of the action.
However, we have also checked the correctness of our results
by computing with a different method a subset of the couplings.
This method consists in identifying the quadratic corrections to 
the linearized equations of motion and integrating them into an action, 
as done in [5].

Some of the final results for the three point correlation functions 
are not new, and can be used as a check on our lengthy calculation and 
procedure. 
In particular, in the super Yang-Mills case (${\rm AdS}_5 \times {\rm S}_5$)
we have reproduced exactly the correlation functions for the CPO
$\langle {\cal O}^s_I(x) {\cal O}^s_J(y) {\cal O}^s_K(z) \rangle$
worked out in [5], and the correlators 
$\langle {\cal O}^s_I(x) {\cal O}^s_J(y) {\cal O}^t_K(z) \rangle$
recently computed in [6] (the relevant bulk coupling are
also given in [7]).
Other correlators are new predictions from the AdS/CFT
correspondence. They could in principle be obtained 
from the CPO correlators
by a systematic use of the 
superconformal algebra [21], though we have not attempted to do so.
For the $d=6$, ${\cal N}=(2,0)$ SCFT (${\rm AdS}_7 \times {\rm S}_4$)
we can compare our results for the CPOs with ref. [9].
Actually, their results differ slightly from ours,
though those authors seem to agree with our findings [22].
All other results are new as new is the case of 
the $d=3$, ${\cal N}= 8$ SCFT (${\rm AdS}_4 \times {\rm S}_7$).

We should mention that in obtaining the values of the correlation
functions for certain extremal cases we have implicitly used an
analytic continuation in the conformal dimensions of the operators [23].
These extremal cases have the property that
the bulk three point couplings of the supergravity source 
fields apparently vanish. However this zero 
is compensated by a divergent integral over AdS to produce a 
finite final result. A recent analysis on these issues has appeared 
also in [24], supporting the use of analytic continuation.

\par\vskip.6cm
\item{\bf A1.} {\bf Scalar Spherical Harmonics}
\vskip.4cm
\par

We describe 
the $n$-sphere of radius $\rho = \bar e^{-1}$ by
${\rm S}_n\equiv\{ z^2=\rho^2 | z \in \Bbb R^{n+1} \}$, 
and use scalar spherical harmonics defined by $Y_I=
{\cal C}_{Ii_1...i_k}x^{i_1}...x^{i_k}$, 
where the coordinates $x^i = \bar e z^i$  live on the unit sphere
and the tensors ${\cal C}_{Ii_1...i_k}$ form
an orthonormal basis of the completely symmetric 
traceless tensors so that 
${\cal C}_{Ii_1...i_k} {\cal C}_{J}{}^{i_1...i_k} = \delta_{IJ}$.
The $Y_I$ are eigenfunctions of the ${\rm S}_n$ Laplacian
$$
\bar\nabla^\alpha\bar\nabla_\alpha Y_I=-\bar e^2 k(k+n-1)Y_I.
\eqno(A1.1)
$$
One has 
$$
\eqalignno{
\int_{S_n} d^n z {\sqrt g} Y_{I_1} Y_{I_2} 
&= z_{I_1} \delta_{I_1I_2},&(A1.2a)\cr
\int_{S_n} d^n z {\sqrt g}  Y_{I_1} Y_{I_2} Y_{I_3} 
&= a_{I_1 I_2 I_3}\langle{\cal C}_{I_1}{\cal C}_{I_2}{\cal C}_{I_3}\rangle,
&(A1.2b)\cr}
$$
where   
$$
\eqalignno{
z_I&= \omega_n {(n-1)!! k!\over{(2k+n-1)!!}}\bar e^{-n},
&(A1.3a)\cr
a_{I_1I_2I_3}&=\omega_n  {(n-1)!!\over{(2\alpha+n-1)!!}}
{k_1!k_2!k_3! \over \alpha_1! \alpha_2! \alpha_3!}\bar e^{-n}.
&(A1.3b)\cr}
$$
$\langle{\cal C}_{I_1}{\cal C}_{I_2}{\cal C}_{I_3}\rangle$ 
denotes the unique $SO(n+1)$ scalar contraction
of three tensors  ${\cal C}_{Ii_1...i_k}$ and
$\omega_n$ is the volume of the unit sphere
$$
\omega_n={2\pi^{n+1\over 2}\over\Gamma({n+1\over2})}.
\eqno(A1.4)
$$
In particular, the $Y_I$ form an orthonormal basis of the Hilbert space of 
scalar functions on ${\rm S}_n$.

\vfill\eject

\centerline{\bf REFERENCES}

\item{[1]} J. Maldacena, 
``The Large $N$ limit of Superconformal Field Theories and Supergravity'', 
Adv. Theor. Math. Phys. {\bf 2} (1998), 231, {\tt hep-th/9711200}.

\item{[2]} S. Gubser, I. Klebanov and A. Polyakov, ``Gauge Theory
Correlators from Non-critical String Theory,'' 
Phys. Lett. {\bf B428} (1998), 105, {\tt hep-th/9802109}.

\item{[3]} E. Witten, ``Anti-de Sitter Space and Holography'',
Adv. Theor. Math. Phys. {\bf 2} (1998), 253, {\tt hep-th/9802150}.

\item{[4]} O. Aharony, S. Gubser, J. Maldacena, H. Ooguri and Y. Oz,
``Large $N$ Field Theories, String Theory and Gravity'',
{\tt hep-th/9905111}.

\item{[5]} S. Lee, S. Minwalla, M. Rangamani and N. Seiberg,
``Three Point Functions of Chiral Primary Operators in $D=4$, $N=4$ SYM 
at Large $N$'', Adv. Theor. Math. Phys. {\bf 2} (1998), 697, 
{\tt hep-th/9806074}.

\item{[6]} G. Arutyunov and S. Frolov,
``Some Cubic Couplings in Type IIB Supergravity on $AdS_5\times S^5$ and 
Three-point Functions in SYM$_4$ at Large $N$'',
{\tt hep-th/9907085}.

\item{[7]} S. Lee,  
``$AdS_5$/$CFT_4$ Four-point Functions of Chiral Primary Operators: 
Cubic Vertices'', 
{\tt hep-th/9907108}.

\item{[8]} E. D'Hoker, D. Freedman, S. Mathur, A. Matusis and L. Rastelli,
``Graviton Exchange and Complete 4-point Functions in the AdS/CFT
Correspondence'',
{\tt hep-th/9903196}.

\item{[9]} R. Corrado, B. Florea and R. McNees, 
``Correlation Functions of Operators and Wilson Surfaces
in the $d=6$, $(0,2)$ Theory in the Large $N$ Limit'',
{\tt hep-th/9902153}.

\item{[10]} F. Bastianelli and R. Zucchini,
``Three Point Functions of Chiral Primary Operators in $d=3$, 
${\cal N}= 8$ and $d=6$, ${\cal N}=(2,0)$ SCFT at Large $N$'',
{\tt hep-th/9907047}.

\item{[11]} P. van Nieuwenhuizen,
``The Complete Mass Spectrum of $d=11$ Supergravity
Compactified on $S^4$ and a General Mass Formula 
for Arbitrary Cosets $M_4$'', Class. Quantum Grav. {\bf 2} (1985) 1.

\item{[12]}  B. Biran, A. Casher, F. Englert, M. Rooman and P. Spindel,
``The Fluctuating Seven Sphere in Eleven Dimensional Supergravity'',
Phys. Lett. {\bf 134B} (1984) 179.

\item{[13]} L. Castellani, R. D'Auria, P. Fr\'e, K. Pilch and 
P. van Nieuwenhuizen, 
``The Bosonic Mass Formula for Freund--Rubin Solutions of
$d=11$ Supergravity on General Coset Manifolds'', 
Class. Quantum Grav. {\bf 1} (1984), 339.

\item{[14]} P. Freund and M. Rubin, 
``Dynamics of Dimensional Reduction'',
Phys. Lett. {\bf B97} (1980), 233.

\item{[15]} H. Kim, L. Romans and P. van Nieuwenhuizen,
``The Mass Spectrum of Chiral N=2 D = 10 Supergravity on $S^5$'',
Phys. Rev. {\bf D32} (1985) 389. 

\item{[16]} G. Dall'Agata, K. Lechner and D. Sorokin, 
``Covariant Actions for the Bosonic Sector of D=10 IIB Supergravity'',
Class. Quant. Grav. {\bf 14} (1997) L195-L198, {\tt hep-th/9707044}.

\item{[17]} G. Arutyunov and S. Frolov, ``Quadratic Action for type IIB
Supergravity on ${\rm AdS}_5\times {\rm S}^5$'', 
J. High Energy Phys. {\bf 08} (1999) 024, 
{\tt hep-th/9811106}.

\item{[18]} F. Bastianelli and R. Zucchini,
``Bosonic Quadratic Actions for 11D Supergravity on 
${\rm AdS}_{7/4} \times {\rm S}_{4/7}$''
Class. Quant. Grav., to be published,
{\tt hep-th/9903161}.

\item{[19]} D. Freedman, S. Mathur, A. Matusis and L. Rastelli,
``Correlation functions in the CFT(d)/AdS(d+1) correspondence'',
Nucl. Phys. {\bf B546} (1999) 96, {\tt hep-th/9804058}. 

\item{[20]} I. Klebanov,
``Absorption by Threebranes and the AdS/CFT Correspondence'',
Talk given at Strings '99, {\tt hep-th/9908165}. 

\item{[21]} P. Howe, E. Sokatchev and P. West, 
``Three Point Functions in $N=4$ Yang-Mills'', 
Phys. Lett. {\bf B444} (1998) 341, {\tt hep-th/9808162}.   

\item{[22]} R. Corrado, private communication.

\item{[23]} H. Liu and A. Tseytlin,
``Dilaton - Fixed Scalar Correlators and AdS$_5$ $\times$ S$^5$ - SYM 
Correspondence'',  {\tt hep-th/9906151}.

\item{[24]} E. D'Hoker, D. Freedman, S. Mathur, A. Matusis, L. Rastelli,
``Extremal Correlators in the AdS/CFT Correspondence'',
{\tt hep-th/9908160}.

\bye

\item{[19]} S. P. de Alwis,
``Coupling of Branes and Normalization of Effective
Actions in String / M Theory'',
Phys. Rev. {\bf D56} (1997), 7963, hep-th/9705139.

\item{[20]} I. Bandos, N. Berkovits and D. Sorokin,
``Duality Symmetric Eleven-Dimensional Supergravity and
its Coupling to M-Branes'',
Nucl. Phys. {\bf B522} (1998), 214, hep-th/9711055. 

\item{[]} D. Z. Freedman, S. D. Mathur, A. Matusis and L. Rastelli, 
as discussed in Freedman's conference lecture at ``Strings 98'', available 
at http://www.itp.ucsb.edu/string98/.

\item{[]} H. Liu and A. A. Tseytlin, ``On Four Point Functions in the 
CFT/AdS Correspondence'',
Phys. Rev. {\bf D59} (1999) 086002, hep-th/9807097.

\item{[]} D. Z. Freedman, S. D. Mathur, A. Matusis and L. Rastelli, 
``Comment on 4-Point Functions in the CFT/AdS Correspondence'',
hep-th/9808006.

\item{[8]}  E. D'Hoker and D.Z. Freedman
``General scalar exchange in $AdS_{d+1}$'',
hep-th/9811257.

\item{[9]} O. Aharony, Y. Oz and Z. Yin, 
``M Theory on ${\rm AdS}_p\times {\rm S}_{11-p}$ and Superconformal Field
Theories'', 
Phys. Lett. {\bf B430} (1998), 87, hep-th/9803051.

\item{[10]} S. Minwalla, ``Particles on {$AdS_{4/7}$} and Primary Operators 
on {$M_{2/5}$}-Brane World Volumes'', 
J. High Energy Phys. {\bf 9810} (1998) 002, hep-th/9803053.

\item{[11]} R.G. Leigh and M. Rozali, 
``The Large  $N$  Limit of the $(2,0)$ Superconformal Field Theory'', 
Phys. Lett. {\bf B431} (1998) 311, hep-th/9803068.

\item{[12]} E. Halyo, 
``Supergravity on {$AdS_{4/7}\times S^{7/4}$} and  M  Branes'', 
J. High Energy Phys. {\bf 04} (1998) 011, hep-th/9803077.

$$
\langle {\cal O}_{I_1}(x) {\cal O}_{I_2}(y) {\cal O}_{I_3}(z) \rangle = 
{\bar R_{I_1I_2I_3}\langle{\cal C}_{I_1}{\cal C}_{I_2}{\cal C}_{I_3}\rangle
\over |x-y|^{\Delta_1+\Delta_2-\Delta_3}
|y-z|^{\Delta_2+\Delta_3-\Delta_1}
|z-x|^{\Delta_3+\Delta_1-\Delta_2}}.
$$

Then, 
$$
\eqalignno{  
\pi_I&=2(2+p)(D-3-p)(ks_I+(k+1+p)t_I),&(.a)\cr    
b_I&=(-1)^p(1+p)(D-2)e^{-1}(s_I-t_I). &(.b)\cr}
$$

$$\eqalignno{
g^{sss}_{I_1I_2I_3}s_{I_1}s_{I_2}s_{I_3}&=
{\over }{\alpha_1 \alpha_2 \alpha_3 \over (k_1+ )(k_2+ )(k_3+ )}

&(.a)\cr

g^{sst}_{I_1I_2I_3}s_{I_1}s_{I_2}t_{I_3}&=
{\over }{\alpha_1 \alpha_2 \alpha_3 \over (k_1+ )(k_2+ )(k_3+ )}

&(.b)\cr

g^{tts}_{I_1I_2I_3}t_{I_1}t_{I_2}s_{I_3}&=
{\over }{\alpha_1 \alpha_2 \alpha_3 \over (k_1+ )(k_2+ )(k_3+ )}

&(.c)\cr

g^{ttt}_{I_1I_2I_3}t_{I_1}t_{I_2}t_{I_3}&=
{\over }{\alpha_1 \alpha_2 \alpha_3 \over (k_1+ )(k_2+ )(k_3+ )}

&(.d)\cr

}
$$

$$\eqalignno{
g^{sss}_{I_1I_2I_3}s_{I_1}s_{I_2}s_{I_3}&=
{\over }{\alpha_1 \alpha_2 \alpha_3 \over (k_1+ )(k_2+ )(k_3+ )}

&(.a)\cr

g^{sst}_{I_1I_2I_3}s_{I_1}s_{I_2}t_{I_3}&=
{\over }{\alpha_1 \alpha_2 \alpha_3 \over (k_1+ )(k_2+ )(k_3+ )}

&(.b)\cr

g^{tts}_{I_1I_2I_3}t_{I_1}t_{I_2}s_{I_3}&=
{\over }{\alpha_1 \alpha_2 \alpha_3 \over (k_1+ )(k_2+ )(k_3+ )}

&(.c)\cr

g^{ttt}_{I_1I_2I_3}t_{I_1}t_{I_2}t_{I_3}&=
{\over }{\alpha_1 \alpha_2 \alpha_3 \over (k_1+ )(k_2+ )(k_3+ )}

&(.d)\cr

}
$$

\bye